\documentclass[prl,twocolumn,amssymb,amsfonts]{revtex4}
\usepackage{epsfig}
\usepackage{amsmath}

\begin{document}

\title{Analysis of Trapped Quantum Degenerate Dipolar Excitons}

\author{Ronen Rapaport, Gang Chen, and Steven Simon}
\affiliation{Bell Laboratories, Lucent Technologies, 600 Mountain
Avenue, Murray Hill, New Jersey 07974}

\begin{abstract}
The dynamics of quantum degenerate two-dimensional dipolar excitons
confined in electrostatic traps is analyzed and compared to recent
experiments. The model results stress the importance of artificial
trapping for achieving and sustaining a quantum degenerate exciton
fluid in such systems and suggest that a long-lived, spatially
uniform, and highly degenerate exciton system was experimentally
produced in those electrostatic traps.
\end{abstract}

\maketitle

Excitons are excellent candidates for producing Bose-Einstein (BE)
quantum statistical effects, such as condensation and superfluidity,
on a semiconductor platform
\cite{SnokeScience2002,EisensteinNature2004}. Observation of such
effects can open up new and exciting opportunities for both
fundamental and applied research. A crucial step in this direction
is the ability to obtain long-lived quantum degenerate excitons,
which practically means sustaining a high density exciton cloud for
times long enough for the excitons to cool down to the lattice
temperature. One of the more promising two-dimensional (2D) exciton
systems for this purpose is the dipolar (or "indirect") exciton
system in double quantum well (DQW) structures. In this system, the
constituent electrons and holes reside in difference quantum wells,
giving the excitons very long recombination lifetimes (in the
microsecond range) which should be much longer than their
thermalization time with the host lattice. In addition to being able
to cool the excitons, to observe BE effects, one must also be able
to obtain high densities. Unfortunately, the combined task of
cooling an exciton fluid while maintaining high density has turned
out to be quite difficult, as recent studies have revealed
\cite{VorosPRL2005,RapaportPRB2006}: the inherent strong dipolar
repulsion between pairs of excitons, while desirable for eliminating
the formation of $eh$ droplets and other complexes, also leads to a
rapid spatial expansion of the initially dense exciton cloud. As a
result, the exciton density quickly drops to below the critical
value for quantum degeneracy. In addition, the strong exciton
expansion prevents the formation of a static fluid and the dipolar
repulsion energy is converted into exciton kinetic energy during the
expansion, which can result in a heating source which slows the
cooling process. This fast expansion of the exciton cloud thus poses
a serious technical problem to producing BE effects in exciton
systems.

One possible solution is to continuously supply cold dipolar
excitons, using long distance in-plane charge transport ("exciton
rings") as a cooling mechanism
\cite{ButovPRL2004,RapaportPRL2004,YangCondmat2006}. Recently we
have been trying to find a solution by implementing spatially
confining potential traps for the dipolar excitons in the plane of
the quantum wells, an approach that is somewhat analogous to that of
atom trapping and cooling. Two different trapping methods have been
attempted experimentally: strain-induced \cite{SnokeSSC2005} and
electrostatic \cite{RapaportPRB2005, ChenPRB2006,ButovCondmat2005}
traps (Xtraps). The Xtraps, the design of which is given in
Ref.~\cite{RapaportPRB2005} seem to be particularly promising as
they are easily tunable in size and depth and have sharp boundaries,
all of which are important for achieving high exciton densities, as
was recently demonstrated experimentally \cite{ChenPRB2006}.

In light of these recent experiments, it is important to analyze and
understand the exciton dynamics in such Xtraps and their critical
role in achieving a long lifetime and spatially uniform degenerate
fluid of excitons. The calculations in this paper are based on the
models developed by Ivanov et. al.
\cite{IvanovPRB1999,IvanovEurophysLett2001}, and extended to include
the Xtrap potential and expansion induced exciton heating. They
explain our previous experimental observations and reveal the
detailed dynamics of the dipolar exciton fluid confined in a trap.
In particular, these calculations show the advantages of this
trapping method, and suggest that the exciton fluid confined in the
Xtraps is highly degenerate and spatially uniform, over long time
periods of the order of the exciton lifetime.

\begin{figure}[htb]
\begin{center}
\includegraphics[scale=0.35]{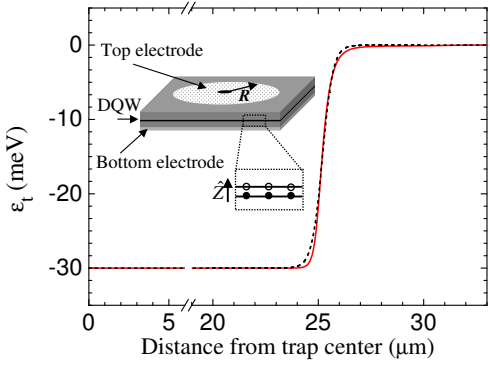}
\caption{The calculated Xtrap potential radial profile (solid
line) and its analytic approximation (dashed line). The inset
shows an schematic diagram of an Xtrap design
geometry.}\label{figure1}
\end{center}
\end{figure}

The inset of Fig.~\ref{figure1} shows an electrostatic trap design
\cite{RapaportPRB2005}, where dipolar excitons (with a dipole moment
given by $\overrightarrow{d_X}=-ez_0\hat{z}$, $z_0$ being the
effective dipole length) are trapped under a local circular
electrostatic gate with radius $R$. When an electrical bias is
applied to the gate, the (negative) dipole-field interaction energy
profile $\varepsilon_{t}$ at the DQW position, plotted in
Fig.~\ref{figure1} (solid line), effectively confines the excitons
to the region under the circular gate with the trapping force given
by $\textbf{F}_t=-\nabla \varepsilon_{t}$ \cite{RapaportPRB2005}.
For calculation convenience, we approximate (Fig.~\ref{figure1},
dashed line) the xtrap potential with an analytic function:
$\varepsilon_{t}(r)=\frac{\varepsilon_t^0}{2}\left(1+tanh\frac{2(r-R)}{\delta}
\right)$. Here, $\varepsilon_t^0$ is the energy at the trap center,
which is proportional to the applied voltage, and $\delta$ is the
effective "thickness" of the boundary region, proportional to the
distance between the the top gate and bottom electrode.

\vspace*{0cm}

Two scattering mechanism for dipolar excitons are included in this
model. The first one is the fast, density dependent exciton-exciton
scattering, which is dominant at high exciton densities. Such a
scattering process, with a characteristic time much smaller than any
other time scale of the dynamics, yields internal exciton
equilibrium  and therefore a well defined exciton temperature,
$T_X$. It is also responsible for a density dependent diffusion
coefficient, $D_{XX}$, calculated by Ivanov et al. in Ref.
\cite{IvanovPRB1999}. The other scattering mechanism is due to
exciton-QW disorder interaction, which is density independent but
yields a diffusion coefficient,$D_{do}$, which depends on the QW
width to the sixth power (see Refs.
\cite{VorosPRL2005,RapaportPRB2006}). We also assume that there is a
single characteristic time ($\tau_l$) for the thermalization of the
exciton gas with the lattice (with a lattice temperature $T_l$). The
dynamics of a dipolar exciton fluid in an Xtrap can then be
described by two coupled equations for the exciton fluid density and
temperature. The first nonlinear diffusion equation describes the
time evolution of the exciton fluid:
\begin{equation}
\frac{\partial n_X}{\partial t}+\nabla\cdot
(\textbf{J}_D+\textbf{J}_d+\textbf{J}_t)+\frac{n_X}{\tau_X}-I_X(r,t)=0.
\label{excitoneq}
\end{equation}

Here, $n_X\equiv n_X(\textbf{r},t)$ is the exciton density profile.
$\textbf{J}_D$, $\textbf{J}_d$, and $\textbf{J}_t$ are the three
exciton currents which are driven by diffusion, dipolar repulsion,
and the trapping potential, respectively. $I_X(r,t)=I_X(r)\delta
(t)$ is the exciton source for a short optical pulse excitation and
$\tau_X$ is the lifetime of the excitons, which is assumed to be
density independent. These exciton currents are related to the
various forces through the exciton mobility
$\textbf{J}=n_X\mu\textbf{F}$. The exciton mobility, $\mu$, is
related to the effective exciton diffusion coefficient
$D=D_{XX}D_{do}/(D_{XX}+D_{do})$, through Einstein relations
 $\mu=(D/ kT_0)(e^{T_0/T_X}-1)$, with $T_0=(2\pi
\hbar^2n_X)/(kgm_X)$ being the degeneracy temperature. Here,
$m_X=0.2m_e$ is the exciton mass and $g=4$ is the exciton spin
degeneracy. The dipole-repulsion force, $\textbf{F}_d=-\alpha \nabla
n_X$ is due to the mean field dipole-dipole interaction energy
\cite{BenTabouPRB2001}: $\varepsilon_{dd}=4\pi
ed_Xn_X/\epsilon\equiv\alpha n_X$, where $\epsilon$ is the
background dielectric constant. The diffusive force is
$\textbf{F}_D=-\nabla \zeta=-\frac{kT_0/n_X}{e^{T_0/T_X}-1}\nabla
n_X$, where $\zeta = kT_Xln\left(1-e^{-T_0/T_X}\right)$ is the
chemical potential in the non-interacting limit. The trapping force
for the dipolar excitons can be derived from the above trapping
potential:
$\textbf{F}_t=-\frac{\varepsilon_t^0}{\delta}\left[1-tanh^2\frac{2(r-R)}{\delta}
\right]\hat{\textbf{r}}$. The three expressions for the exciton
currents in Eq.~(\ref{excitoneq}) are therefore: (a)
$\textbf{J}_d=-\mu\alpha n_X\nabla n_X$, (b) $\textbf{J}_D=-D\nabla
n_X$, and (c) $\textbf{J}_t=-\mu
n_X\frac{\varepsilon_t^0}{\delta}\left[1-tanh^2\frac{2(r-R)}{\delta}\right]\hat
{\textbf{r}}$.

The second equation describes the time evolution of the exciton
fluid temperature, $T_X$:
\begin{equation}
\frac{dT_X}{dt}=-\left[\frac{1}{k_BN_X}\left(\frac{\partial
E_{dd}}{\partial t}\right)_{N_X}+\frac{T_X-T_l}{\tau_{l}}\right],
\label{excitonTeq}
\end{equation}
where $E_{dd}$, the total potential energy due to the dipole-dipole
interactions is given by: $E_{dd}=\int\alpha
n_X^2(\textbf{r},t)d^2r$ and $N_X(t)=\int n_X(\textbf{r},t)d^2r$ is
the total number of dipolar excitons. The first term on the RHS of
Eq. \ref{excitonTeq} represents the heating of the exciton gas due
to the driven expansion and the last term represents the
thermalization of the excitons with the lattice.

Eqs.(\ref{excitoneq}),(\ref{excitonTeq}) are coupled and can be
numerically evaluated. For the following calculations, we use
parameters that are typical for our experiments with GaAs double QW
structures \cite{ChenPRB2006}. The size of the trap is taken as
$R=25~\mu m$, consistent with the experiments performed in
\cite{ChenPRB2006}. The depth of the trap is fixed at
$\varepsilon_t^0=30~meV$. For a short ($\sim1ns$) pulse excitation
with a gaussian profile, we assume that dipolar excitons are
subsequently created at the center of the trap with an initial
gaussian profile: $n_X(r,t=0)=n_0 e^{-r^2/w_0^2}$ where typically
for our experiments, $w_0=15~\mu m$ and
$n_0=1.5\times10^{11}cm^{-2}$. The initial exciton temperature is
taken to be the energy difference between the optical excitation
energy and the dipolar exciton energy, $T_X(t=0)\simeq50K$, which is
much hotter than the lattice temperature $T_l=1.4K$, as we assume
that the optically excited carriers rapidly transform into hot
dipolar excitons. The exciton lifetime, $\tau_X$ is position
dependent. Inside the trap, the excitons are dipolar and their
lifetime is in microsecond range. For the particular trap depth
$30~meV$, we use a typical value of $\tau_X=\tau_{X_{ID}}=1.5\mu s$.
Outside the trap, the excitons are direct and their lifetime is
taken as $\tau_X=\tau_{X_{D}}=0.1ns$. In the trap boundary region,
the dipolar excitons are subject to ionization due to the in-plane
fringing fields that depends on the geometry of the trap design (see
Ref.~\cite{RapaportPRB2005} for details). This ionization process
yields an effective boundary lifetime for the excitons
($\tau_{trap}$). We have shown analytically \cite{RapaportPRB2005}
that this lifetime, and thus the right Xtrap design, is crucial for
the effectiveness of the Xtrap, as was further confirmed by our
experiments \cite{ChenPRB2006} and the following calculations.

\vspace*{0cm}
\begin{figure}[htb]
\begin{center}
\includegraphics[scale=0.35]{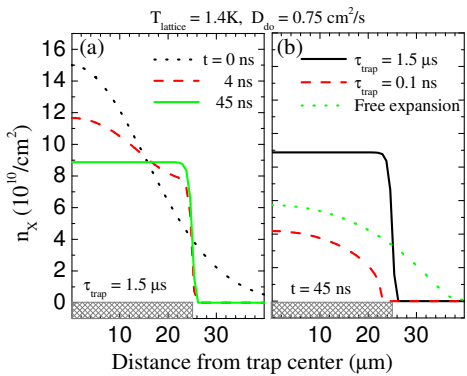}
\caption{Radial profile of the exciton density distribution for
(a) a high quality Xtrap ($\tau_{trap}=1.5\mu s$) at various
times, and (b) High quality Xtrap ($\tau_{trap}=1.5 \mu s$, solid
line), low quality Xtrap ($\tau_{trap}=0.1ns$, dashed line), and
free expanding excitons (dotted line) at $t=45ns$. The bar at the
bottom of each figure mark the radial extent of the
Xtrap.}\label{figure2}
\end{center}
\end{figure}

Fig.~\ref{figure2}a plots the radial density profile of the
excitons in a high quality Xtrap with negligible boundary
ionization, in which case the boundary lifetime is only limited by
the intrinsic lifetime of the excitons ($\tau_{trap}=\tau_X$). The
values of the diffusion coefficients are chosen by fitting the
expansion dynamics (the cloud FWHM as a function of time in
particular) of a free expanding excitons to experimental results
(see Ref.~\cite{RapaportPRB2006}). In a good agreement with the
experimental results of Ref.~\cite{ChenPRB2006}, within tens of
nanoseconds the profile flattens, as the driven expansion by the
dipole repulsion pushes the excitons to the reflecting trap
boundary. These two competing forces result in a stable flat
exciton density profile (the density profile with the lowest total
energy) that decays with the characteristic radiative lifetime of
the excitons, $\tau_X$. As we will show below, as the profile
flattens out, the exciton quickly cools to the lattice temperature
and becomes highly degenerate. This degeneracy is maintained on
the time scale of the exciton radiative lifetime.

This result is strikingly different from either a free expanding
exciton cloud or excitons in a low quality Xtrap with a considerable
boundary ionization, as depicted in Fig.~\ref{figure2}b. In the case
of free expanding excitons, the density rapidly decreases, as the
driven expansion continues to spread the exciton cloud to large
radii. For a low quality trap ($\tau_{trap}=0.1ns$), excitons
reaching the trap boundaries are effectively eliminated by
ionization, rapidly depleting the expanding exciton cloud. Note that
in this case, the exciton profile remains curved and does not
flattens like the exciton pool in the high quality trap. This is
also in agreement with the experimental profile found for a low
quality Xtrap in Ref.~\onlinecite{ChenPRB2006}.

\vspace*{0cm}
\begin{figure}[htb]
\begin{center}
\includegraphics[scale=0.35]{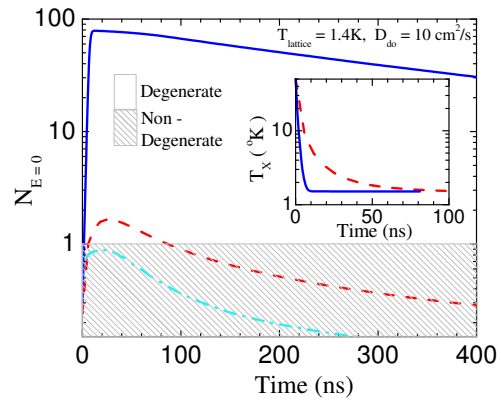}
\caption{The exciton ground state occupancy, $N_{E=0}$, as a
function of time for a high quality Xtrap ($\tau_{trap}=1.5\mu s$,
solid line), a low quality trap ($\tau_{trap}=1ns$, dotted line)
and free expanding excitons (dashed line). The inset shows the
corresponding exciton temperature as a function of time for the
high quality Xtrap (solid line) and free expending excitons
(dotted-dashed line).}\label{figure3}
\end{center}
\end{figure}

To see the dramatic effect of trapping on high mobility excitons
($D_{do}= 10cm^2/s$), we show the calculated ground state occupation
number of the 2D excitons, $N_{E=0}=(\exp(T_0/T_X)-1)$, as a
function of time after photoexcitation, for excitons in a high and
low quality Xtrap as well as for free excitons. Initially, the
ground state occupation is much smaller than unity, as the excitons
are hot and essentially classical. The exciton cloud then start to
expand and cool. As the transition to a distinctive BE statistics
arises when $T_0/T_X\sim 1$, there is a competition between cooling,
that tend to increase $N_{E=0}$ and expansion that tend to decrease
it due to the drop of $n_X$. In a high quality Xtrap, expansion is
limited to the Xtrap boundaries, and thus reaches steady state after
$\sim 10$ nanoseconds. Cooling is then efficient (as is seen in the
inset of Fig.~\ref{figure3}) and the excitons reach a stable high
degenerate state with $N_{E=0}\gg 1$, that decays slowly on the time
scale of $\tau_X$. On the contrary, if the excitons are freely
expanding (no trap), their density continues to drop as they expand.
While they continue to cool by interaction with the lattice, the
expansion itself heats up the cloud, as more internal potential
energy is lost and converted into heat by the fast exciton-exciton
interaction. This leads to a slower net cooling rate for the free
expanding excitons compared with the trapped ones, and consequently
the excitons reach only marginal quantum degeneracy before their
density drops too low due to the fast driven expansion. Finally, in
a low quality trap, the drop in exciton density due to boundary
ionization dominates over the cooling, completely preventing the
excitons from reaching degeneracy.

While those calculations strongly suggest that the exciton pool that
was created in Ref.~\cite{ChenPRB2006} is highly degenerate for
hundreds of nanoseconds, it is based on the assumption that the
cooling rate is indeed much shorter than the exciton lifetime. While
this is a very reasonable assumption (see Ref.
\cite{IvanovPRB1999}), a clear experimental evidence of this
degeneracy that is independent on model assumptions is still
essential, and experiments looking for such conclusive evidence are
ongoing.

In summary, the dynamics of two-dimensional dipolar excitons
confined in electrostatic traps is modeled and agrees well with
recent experiments. The model shows that artificial trapping is
crucial for achieving and sustaining a quantum degenerate exciton
fluid is such systems and suggest that a spatially uniform, highly
degenerate exciton pool was experimentally produced in those
electrostatic traps.

%-----------------------------------------------------------------------------

\end{document}